\documentclass[smallextended]{svjour3}       
\smartqed  
\usepackage{graphicx}
\usepackage{multirow}
\usepackage{tabularx}
\usepackage{soul}
\usepackage{float}
\usepackage{amsmath}
\usepackage{amsfonts}
\usepackage{lineno,hyperref}
\def\makeheadbox{\relax}
\begin{document}

\title{Privacy-Aware Recommender Systems Challenge on Twitter's Home Timeline
}


\author{
Luca Belli 
\and
Sofia Ira Ktena 
\and
Alykhan Tejani \and
Alexandre Lung-Yut-Fong \and
Frank Portman \and
Xiao Zhu \and
Yuanpu Xie \and
Akshay Gupta \and
Michael Bronstein \and
Amra Deli\'{c}\and
Gabriele Sottocornola \and
Vito Walter Anelli\and
Nazareno Andrade \and
Bart P. Knijnenburg\and
Jessie Smith \and
Wenzhe Shi
}

\authorrunning{Belli et al.} 

\institute{Luca Belli, \at Twitter \\
              \email{lb@twitter.com}           
        \and
            Sofia Ira Ktena \at Deepmind Technologies (work beformed while at Twitter)
        \and
            Alykhan Tejani, \at Twitter
        \and
            Alexandre Lung-Yut-Fong, \at Twitter 
        \and
            Frank Portman, \at Twitter 
        \and
            Xiao Zhu, \at Twitter 
        \and
            Yuanpu Xie, \at Twitter 
        \and
            Akshay Gupta, \at Twitter 
        \and
            Michael Bronstein, \at Twitter 
        \and
            Amra Deli\'{c}\footnote{Work performed while at TU Wien, Vienna, Austria}, \at University of Sarajevo, Sarajevo, B\&H 
        \and
            Gabriele Sottocornola, \at Free University of Bozen-Bolzano, Italy
        \and
            Vito Walter Anelli, \at Politecnico di Bari, Italy
        \and
            Nazareno Andrade, \at Universidade Federal de Campina Grande, Brazil
        \and
            Bart P. Knijnenburg, \at Clemson University, United States
        \and
            Jessie Smith, \at University of Colorado Boulder, United States
        \and
            Wenzhe Shi, \at Twitter 
}

\date{}

\maketitle

\begin{abstract}
 Recommender systems constitute the core engine of most social network platforms nowadays, aiming to maximize user satisfaction along with other key business objectives. Twitter is no exception. Despite the fact that Twitter data has been extensively used to understand socioeconomic and political phenomena and user behaviour, the implicit feedback provided by users on Tweets through their engagements on the Home Timeline has only been explored to a limited extent. At the same time, there is a lack of large-scale public social network datasets that would enable the scientific community to both benchmark and build more powerful and comprehensive models that tailor content to user interests. By releasing an original dataset of 160 million Tweets along with engagement information, Twitter aims to address exactly that. During this release, special attention is paid to maintaining the dataset in sync with the Twitter platform. Apart from user privacy, this paper touches on the key challenges faced by researchers and professionals striving to predict user engagements. It further describes the key aspects of the RecSys 2020 Challenge that was organized by ACM RecSys in partnership with Twitter using this dataset.
\end{abstract}
\keywords{privacy aware recommender systems \and large-scale dataset \and personalization \and engagement prediction  
}

\section{Introduction}
Twitter is what's happening around the world. The platform strives to keep users informed with relevant and healthy content at a global level. In the context of online information overload, it is extremely important for both producers of Tweets to reach the right (target) audience, and for consumers to be recommended the most relevant content (normally generated by the hundreds or thousands of people they follow). 
Twitter's Home timeline, the default starting point for most Twitter users, displays a stream of Tweets from accounts the user has chosen to follow on Twitter. Users can decide if they want Tweets to be displayed in a reverse chronological order, or if they want them algorithmically ranked. In the latter case, every Tweet is scored, with the score provided by a predictive model and indicating how interesting and engaging the content would be for the user.   

Over the years, Twitter has been, and continues to be, a great motivation and source for a number of research works studying user behavior in social media platforms, and the effect of these platforms on the society as a whole. Some of the most prominent works analyzed topics such as the role of Twitter in the social communication \cite{murthy2018}, how to use a global ``mood'' on Twitter to predict the stock market movements \cite{bollen2011}, how to detect events \cite{weng2011}, or influenza pandemics by using content available on Twitter \cite{aramaki2011}, or even analyzing mental health issues of Twitter users \cite{coppersmith2014,odea2015}. 

However, it is not so often that we see work addressing or analyzing challenges of the core Twitter task, i.e., delivering relevant content to users. For instance, back in 2010~\cite{duan2010}, researchers introduced a learning to rank method to distinguish relevant from irrelevant Tweets that integrated information about the authority of the Tweet creator (``publisher''), and features of the Tweet. In \cite{chen2012}, a collaborative ranking method was proposed. The method accounted for user historical preferences of the content (i.e., Tweets), social relationships between users, as well as authority of the Tweet creator, and the quality of the Tweet content. Then, in~\cite{demaio2019}, a deep learning ranking algorithm that incorporated a time dimension, in order to rank tweets accounting for the time of day when a user is active on Twitter was evaluated. Earlier work~\cite{pennacchiotti2012} suggested a recommendation approach based on similarity between active user's Tweets and their friends' Tweets. With the Home timeline being one of the core products on the platform, the quality of the ranking model is of extreme importance, as it determines the quality of the user experience. People have more conversations and are more likely to come back to the platform when the timeline is optimized to show the most relevant Tweets first. Quality itself is a very personal concept that is hard to define. In this work, engagement is considered as a proxy for quality, i.e., the user interacts with the content if they value it. The goal of this paper is to invite and facilitate a broader research community to explore, within the scope of their interest, the task of delivering / recommending / ranking content such as Tweets, with all its peculiarities.

The task falls within the scope of engagement prediction / user satisfaction, which is omnipresent in recommender systems~\cite{Knijnenburg2012Explaining}. However, the feedback that is received from users on the displayed content is only implicit (as users do not exclusively rate Tweet relevance on their timeline). The lack of explicit feedback makes this an even harder task. In an effort to: a)~address the lack of large public datasets for user engagement prediction, and b)~advance the state-of-the-art in user recommendations with implicit feedback, we release a public dataset of 160 million samples from Twitter's Home timeline, split almost equally between positive and negative examples. To the best of our knowledge, this is the largest public dataset released by a social network platform. The dataset is shared with the RecSys community in the form of a challenge~\footnote{\url{http://www.recsyschallenge.com/2020/}}, paying special attention to user privacy. The release is compliant with existing privacy laws, since a)~the dataset is only includes public Tweets, and b)~any Tweets that are removed by users on the platform are removed from the dataset shortly thereafter. As a result, the size of the training dataset shrinks over time compared to the original release. In the context of the challenge, four different types of interaction (engagement) are considered: Like, Reply, Retweet and Quote Tweet. These interactions are described more elaborately in Section \ref{problem-def}.

The main contributions of the paper are:
\begin{itemize}
    \item A concise \textit{problem definition} of delivering, recommending or ranking Tweets as the engagement prediction, i.e., a binary classification or a ranking problem.
    \item A \textit{set of challenges}, general and specific ones for the presented task. 
    \item A \textit{set of state-of-the-art approaches}, serving as a starting point for the future research endeavors in dealing with the defined task.
    \item And finally, a \textit{detailed description of the publicly released dataset}.
\end{itemize}

The rest of the paper is organized as follows, in section \ref{problem-def}, the problem is defined, section \ref{key-challenges} introduces the key challenges, section \ref{sota} provides an overview of the state of the art approaches, while in section \ref{challenge}, we describe the RecSys Challenge 2020 as an instrument for inviting the research community to participate in this compelling task facilitated by the publicly released large-scale dataset. The paper is concluded with a summary and the future directions in section \ref{conc}. 


\section{Problem definition} \label{problem-def}
Recommender systems optimize for different objectives in different contexts. In an online marketplace, for example, the number of product views or clicks might be the target variable, while in display advertising, conversions might be considered. In many of these cases, the recommender system does not directly optimize for the business objective, e.g., revenue or user retention, but rather a proxy metric like the ones mentioned above. At Twitter, we are mostly interested in engagements on the Home timeline, which is where users see a stream of Tweets from accounts they have chosen to follow, as well as content produced by users outside of their immediate network that is considered relevant to their interests. 
\\

In general, engagement prediction can be formulated either as a binary classification problem (i.e., will the user engage with the content or not), or as a ranking problem (e.g., is the user more likely to engage with this content in comparison to other content candidates). In the former case, predictions will be pointwise, meaning that each candidate will have its own score, normally ranging between 0 and 1 in order to correspond to the probability of engagement. In the latter case, the approaches can be pointwise, pairwise or listwise \cite{liu2009learning}. \\

Let $\mathcal{U}$ be a set of users and $\mathcal{I}$ a set of items. For each user $u \in \mathcal{U}$ we aim to discover  a total ordering over $\mathcal{I}$, where $i \succ_u i'$ implies that $i$ is preferred to $i'$ for $u$. The goal is to learn a ranking function $f$, defined such that
$f : \mathcal{U} \times \mathcal{I} \to \mathbb{R}$ preserves the preference order as much as possible. That is, given a user $u$, for all $i \succ_u i'$, we want $f$ to satisfy $f(u, i) \succ_u f(u, i')$.

In pointwise approaches, each item is assumed to have an ordinal score. Ranking can, then, be formulated as a regression problem in which the absolute value of each item is estimated as an absolute quantity. Such techniques do not consider the interdependency across items. In pairwise approaches, the ranked list is decomposed into a set of item pairs. Ranking is, therefore, considered as the classification of pairs of items, such that the classifier is trained by minimizing the number of misorderings in ranking. Listwise approaches take the entire ranked list of items for each query as a training instance. As a direct consequence, these approaches are able to differentiate items from different queries, and consider their position in the output ranked list at the training stage. 

\section{Key Challenges}\label{key-challenges}
Recommending the most relevant Tweet for the user's timeline turns out to be a difficult problem to tackle at scale. This section summarizes the main challenges that must be addressed in this endeavour. 

\subsection{Sampling} Every day, hundreds of millions of users log in to Twitter to engage with the existing content or to create new content. Using the total number of Tweets that have been created since the launch of the platform\footnote{https://twitter.com/jack/status/20} would be intractable and massively expensive computationally, therefore modelers need to consider their sampling strategy carefully. As an example, it is often reasonable to sample candidates from the most recent past (limited to a fixed time window).

\subsection{Label Imbalance} Users tend to engage with only a fraction of the Tweets displayed on their timeline. This translates to a problem of class imbalance, especially when there is no negative downsampling performed as part of the training pipeline.

\subsection{Social Graph} The social follow graph, i.e., the graph that contains the information about which user follows whom, provides very valuable contextual features for the engagement prediction task at hand. Previous work on smaller datasets has demonstrated performance gains by leveraging this graph structure between users~\cite{monti2017geometric}. Given the hundreds of millions of users that are active on Twitter, such approaches are not as straightforward or even feasible to adopt. Some users might like a certain author more than others, and storing such information makes the problem quadratic in the number of users.

\subsection{Language} The language on Twitter is much less formal and loosely defined. It is not uncommon for users to Tweet in multiple languages, sometimes within the same Tweet. This makes the use of pre-trained language models, such as Word2Vec~\cite{mikolov2013distributed,mikolov2013efficient,mikolov2013linguistic} and BERT~\cite{devlin2018bert}, more challenging.
Even the use of hashtags (used to categorize a Tweet by topic) might be difficult to interpret and process. As an example, consider hashtags created by concatenating multiple words.

\subsection{Data Shift} The conversation on Twitter can change rapidly. Novel hashtags might be trending as a response to real-world events, or the same ones might mean different things at different times. Trained models might become stale very quickly. One way that this problem can be mitigated is described in \cite{Shiebler2018FightingRA}: by introducing embeddings (e.g., at the user or content level) that are trained more often than the rest of the models.

\subsection{Engineering considerations} Finally we must consider the trade-offs between model capacity (to what extent the model is able to correctly predict the preferences of all the users) and model size, which increases resources, utilization and latency. Given the real time nature of Twitter, the speed of prediction is a key factor for any production model. 

\subsection{Metrics vs Intrinsic Value} It is also worth noting that the (personal) intrinsic value of a recommendation and the metric used to measure it might diverge. While engagement is the main metric used in industry, it might not fully represent the quality of engagement~\cite{Knijnenburg2012Explaining}. One example of this would be people replying to very polarized content with inflammatory comments to express their disagreement. While the engagement metric went up, the user probably found negative value in the recommendation.

\section{State-of-the-art}\label{sota}
We are now going to describe some of the state of the art techniques that have been adopted for recommendation problems.

\subsection{CTR model architectures}
The Neural Collaborative Filtering (CF) model for implicit feedback (only available feedback is engagement) was proposed in~\cite{he2017neural}. Each user and item is initially represented as a sparse input and embedded to a latent representation. This is achieved via a fully connected layer that projects the sparse representation to a dense vector. The user embedding and item embedding are then fed to a multi-layer neural architecture, to map the latent vectors to prediction scores. Each layer of the neural CF layers can be customized to discover certain latent structures of user–item interactions. The final output layer is the predicted score and a standard log loss is used for the optimization. Even though this work claims that content features can be used to represent users and items to address the cold-start problem, in fact, only their corresponding identities are used as input features in the form of one-hot encodings.

The Wide and Deep model~\cite{cheng2016wide} consists of a wide component and a deep component. The former is a generalized linear model that handles cross-product transformations / interactions between binary features. The deep component is a feed forward neural network that handles sparse, high-dimensional categorical features, by first embedding them into a low-dimensional and dense real-valued vector (of dimensionality O(10) to O(100)), and concatenates those with the continuous features. Continuous real-valued features are normalized to $[0, 1]$ by mapping a feature value $x$ to its cumulative distribution function $P(X \leq x)$, divided into $n_q$ quantiles. The normalized value is $\frac{i-1}{n_q-1}$ for values in the $i^{th}$ quantile. Quantile boundaries are computed during data generation.

The Deep FM model~\cite{guo2017deepfm} emphasizes both low- and high-order feature interactions by combining the power of factorization machines (FMs) for recommendations and deep learning for feature learning. In other words, this model consists of an FM component and a deep component. The FM component is described by:

\begin{equation*}
    y_{FM} = \langle w, x \rangle + \sum_{j_1=1}^d \sum_{j_2=j_1+1}^d \langle V_i, V_j \rangle x_{j_1} \cdot x_{j_2}
\end{equation*}

where $d$ is the dimensionality of the input vector, while $x_{j_1}$, $x_{j_2}$ are the vector representations of fields $j_1$ and $j_2$, respectively (field here corresponds to a categorical or continuous variable). $V_i$ and $V_j \in \mathbb{R}$ are latent factors that allow the model to learn a representation whenever $i$ or $j$ appear in the data record, removing the constraint that both features $i$ and $j$ need to be present to train the parameter of their interaction. In the above equation, $\langle w, x \rangle$ reflects the importance of order-1 features, while the second term represents the impact of order-2 feature interactions.

The deep component is a standard feed-forward neural network. In this network structure, the embeddings of the different fields/categories are all of the same size $k$. Furthermore, the latent feature vectors ($V$) serve as learned network weights and are used to compress the input field vectors into the embedding vectors. It is worth highlighting that the FM and deep components share the same feature embedding, which brings two important benefits: 1)~it learns both low- and high-order feature interactions from raw features; 2)~there is no need for expertise feature engineering of the input, as required in Wide \& Deep model.

The Deep \& Cross network~\cite{wang2017deep} explicitly applies feature crossings at each layer and learns highly non-linear interactions of bounded degrees. In contrast to the wide-and-deep model, which hinges on a proper choice of cross features, this approach does not require manual feature engineering and has low computational cost. Similar to~\cite{naumov2019deep}, an embedding is obtained for each category, where a category can be e.g., a country. At each layer $x_{l+1}$, feature crossing is guaranteed by multiplying its input $x_l$ with the original feature vector $x_0$, leading to a highest polynomial degree of $l+1$ for an $l$-layer cross network. At the last layer, the output of the cross network and a deep network are concatenated and a standard log loss with regularisation is used. The cross network shares the spirit of parameter sharing as the factorization machine model and further extends it to a deeper structure, while the number of parameters in a cross network only grows linearly with the input dimension.

xDeepFM~\cite{lian2018xdeepfm} can learn certain bounded-degree feature interactions explicitly, while implicitly learning arbitrary low- and high-order feature interactions. The embedding layer in this model operates in a similar manner with the Deep \& Cross and DeepFM models, in the sense that each field is embedded in a vector of the same dimensionality. One observation made by~\cite{lian2018xdeepfm} is that the Deep \& Cross network learns a special type of high-order feature interactions, where each hidden layer is a scalar multiple of $x_0$. On the contrary, in each layer of xDeepFM higher order interactions are computed using the Hadamard product between $X_0$ and $X_l$. In particular, for hidden layer $l$ with $H_l$ feature vectors, the output is calculated as:

\begin{equation*}
    X_{h,*}^l = \sum_{i=1}^{H_{l-1}}\sum_{j=1}^m W_{ij}^{k,l}\big(X_{i,*}^{l-1} \circ X_{j,*}^0 \big)
\end{equation*}

where $m$ is the number of fields/categories (e.g., user id, interests, gender etc.), $1 \leq h \leq H_l$ and $W^{k,l} \in \mathbb{R}^{H_{l-1} \times m}$ is the parameter matrix for the $h$-th feature vector. Hence, the output of the $l$-th layer is also a matrix $X_l \in \mathbb{R}^{H_l \times D}$, with $D$ being the dimensionality of the field embedding.

In the DLRM system~\cite{naumov2019deep}, each categorical feature is represented by an embedding vector of the same dimension $D$, as previously done in Deep \& Cross and xDeepFM models. Unlike other architectures, the continuous features are transformed by a multi-layer perceptron (MLP), which yields a dense representation of the same length as the embedding vectors.

The model also computes second-order interactions of different features explicitly, following the intuition for handling sparse data in FMs~\cite{koren2009matrix}, optionally passing them through MLPs. This is done by computing the dot product between all pairs of embedding vectors and processed dense features. These dot products are concatenated with the original processed dense features and post-processed with another MLP (the top or output MLP), and fed into a sigmoid function to yield a click probability. DLRM interacts embeddings in a structured way that mimics factorization machines to significantly reduce the dimensionality of the model by only considering cross-terms produced by the dot-product between pairs of embeddings in the final MLP.

\subsection{Hashing}
In \cite{chapelle2015simple}, feature hashing to regulate the size of the CTR model is used. The idea behind hashing is to reduce the number of values a feature can take by projecting it to a lower dimensional space. This is a commonly used strategy for ID features and there are two major strategies for hashing. The first approach hashes each feature $f_i$ into a $d_{f_i}$ dimensional space and concatenate the codes, resulting in $\sum_i d_{f_i}$. The alternative approach hashes all features into the same space, when a different hash function can be used for each feature.

A collision between two frequent values can lead to a degradation in the log likelihood. An interesting point made in~\cite{chapelle2015simple} is regarding using multiple hash functions. This is to alleviate the potential issue of degradation, in a similar manner that the Bloom filter operates~\cite{bloom1970space}, by replicating each value using different hash functions.

\subsection{Handling continuous features}
Normalization is considered an important step for continuous features and the approach used in~\cite{cheng2016wide} mapped features to the $[0,1]$ range by splitting their cumulative distribution function into $n_q$ quantiles. The quantile boundaries are computed during data generation. \cite{chapelle2015simple} also uses quantization of the continuous features before feeding them to the prediction model. Even though it is not exactly a normalization approach, Facebook's DLRM transforms the continuous features using a multi-layer perceptron (MLP), in order to yield a dense representation of the same length as the embedding vectors used for the categorical features.
\section{RecSys 2020 Challenge}\label{challenge}
Twitter has partnered with ACM RecSys to sponsor the 2020 challenge. 
The task of the challenge is user engagement prediction, as described in Section \ref{problem-def}.
As part of the challenge, participants are invited to train a model on the data that is publicly released and to beat the baseline that is made available.

\subsection{Dataset description}
An engagement dataset is openly released\footnote{\url{http://recsys-twitter.com/}}. The dataset comprises of (approximately) 160 million possible engagements sampled over one week. Another 40 million are sampled from the following week and split evenly in half for validation and testing.

\subsubsection{Input features}
The dataset features are described in detail in Table~\ref{tab:feature_list}.

The features are divided into three separate feature groups: \textit{user-}, \textit{Tweet-} and  \textit{engagement features}. There are two instantiations of \textit{user features}, one for the author (producer) and one for the reader (consumer) of the Tweet. \textit{Tweet features} group all the attributes describing the original Tweet, that is possibly engaged with by the consumer. Finally, the \textit{engagement features} contain all the details of the engagement itself.

\begin{table}[H]
\resizebox{\linewidth}{!}{%
\begin{tabular}{|c| l | l | l |}
\hline
~ & \multicolumn{1}{c|}{\textbf{Feature name}} & \multicolumn{1}{c|}{\textbf{Feature type} } & \multicolumn{1}{c|}{\textbf{Feature description} } \\
 \hline
{\multirow{5}{*}{\textbf{User features}}} & userId & \textit{string}   & User identifier (hashed) \\
                                     & follower count & \textit{int}                  & Number of followers of the user                                  \\
                                      & following count & \textit{int}                  & Number of accounts this user is following                        \\
                                       & is verified & \textit{bool}                      & Is the account verified?                                         \\
                                    &  account creation timestamp in ms & \textit{int}  & Unix timestamp (in seconds) of the creation time of the account \\ 
  \hline
{\multirow{7}{*}{\textbf{Tweet features}}} &  tweetId & \textit{string}                            & Tweet identifier (hashed)                                                 \\
                                     &  presentMedia & \textit{list[string]}                  & Tab-separated list of media types; media type can be in (Photo, Video, Gif)                                  \\
                                      &  presentLinks & \textit{list[string]}                  & Tab-separated list of links included in the tweet (hashed)                       \\
                                          &  presentDomains & \textit{list[string]}                  & Tab-separated list of domains (e.g. twitter.com) included in the tweet (hashed)   
                           \\           &  tweetType & \textit{string}                     & Tweet type, can be either Retweet, Quote, Reply, or Toplevel                                         \\
                                       & language & \textit{string}  & Identifier corresponding to inferred language of the tweet \\
                                       &  tweet timestamp& \textit{int}                      & Unix timestamp, in seconds of the creation time of the Tweet                                         \\
                                       &  tweet tokens & \textit{list[int]}   & Ordered list of Bert ids corresponding to Bert tokenization of Tweet text \\
                                            &  tweet hashtags & \textit{list[string]}   & Tab-separated list of hashtags present in the tweet \\
\hline                                      
\multirow{5}{*}{\textbf{Engagement features}} & reply engagement timestamp & \textit{int}  & Unix timestamp, in seconds, of the Reply engagement if one exists  \\
 & retweet engagement timestamp & \textit{int}  & Unix timestamp, in seconds, of the Retweet engagement if one exists                                  \\
 & quote engagement timestamp & \textit{int}  & Unix timestamp, in seconds, of the Quote engagement if one exists                     \\
  & like engagement timestamp & \textit{int}  & Unix timestamp, in seconds, of the Like engagement if one exists     
  \\
  
  & engageeFollowsEngager    & \textit{bool}                  & Does the account of the engaged tweet author follow the account that                                         \\
                                      &  &                     & has made the engagement?                                         \\ \hline
\end{tabular}}
\caption{List of features provided for the challenge dataset}
\label{tab:feature_list}
\end{table}

\subsubsection{User privacy}
Previous dataset releases have disclosed private information in a anonymized format. However, user- and item anonymization has proven ineffective to linkage attacks (e.g., \cite{Sweeney1997GuaranteeingAW} and \cite{Narayanan2006HowTB}) that de-anonymize the data by joining with publicly available datasets on seemingly innocuous features. 

In contrast, this dataset contains public Tweets that are accessible via the Twitter API\footnote{https://developer.twitter.com/}. No private information is disclosed. To further increase the difficulty of misusing the dataset, we took extra steps described in following sections.
The goal was to provide a dataset that is useful and stimulating for researchers, while at the same time preventing anyone from learning private information about users.

\paragraph{Developer policy and IRB} The use of the dataset is subject to Twitter's Developer Policy\footnote{https://developer.twitter.com/en/developer-terms/policy}, which, among other things, restricts ``off-Twitter matching'' to data that has been directly provided by the person and/or public data. This would prevent researchers from using the dataset to conduct inference attacks on private datasets.

We also encourage researchers to discuss any use of the dataset outside the context of the RecSys Challenge with their local ethics review board, if available. For instance, we note that U.S. Academic and Government researchers would be required to apply for Institutional Review Board approval to use the dataset. Most uses of the dataset would qualify for ``exempt review'' under Category 4 of federal IRB guidelines (``Secondary research for which consent is not required: Secondary research uses of identifiable private information or identifiable biospecimens'').

\paragraph{Creating pseudo-negative features} For public profiles, all Tweet engagements are public. This made it easy for us to create the first half of the dataset, i.e., the positive examples of engagements. We also wanted to give examples of negative interactions (i.e., this user did \emph{not} engage with this item), but disclosing this information will create a privacy leak. Negative examples are items the user might have seen but not engaged with. However, a set of such examples would reveal what content was seen by users---this is private information. To get around this, we created the pseudo-negative dataset as follows: for each user we considered all the Tweets that were created by their followers in the considered timeframe and removed the positive examples (i.e., the Tweets that were engaged with). We sampled from the set of remaining Tweets, which does not distinguish between negative examples (items the user saw and did not engage with) and items the user did not see, thereby effectively protecting this private information.

\paragraph{Scrubbing deleted content}
\label{gdpr}A novel challenge we encountered in the creation of the dataset was to keep it continuously synced with the Twitter platform. This means that if a user deletes a Tweet and/or their profile (or makes it private), this has to be reflected in the dataset. While we acknowledge that a shared and static dataset is fundamental for the reproducibility of the research, we wanted to prioritize user's choice. 

The way we are solving this problem is by keeping the dataset on the website constantly updated. A change in the system will be promptly reflected in the publicly available dataset. The challenge competitors are required to make the necessary changes in the data they have already downloaded in order to keep them compliant. This requirement is also reflected in Twitter's Developer Policy. To facilitate the task, we are also going to provide a list of the deleted user/Tweet ids.
This process makes the whole dataset dynamic (including the validation and test set, used for scoring). Given the size of the dataset, it is very unlikely that the scrubbing will results in a meaningful reduction in dataset size.

\paragraph{Handling text features}
Since providing raw Tweet text could make the reconstruction of the dataset immediate, we tokenized the text and provided the IDs of each token according to BERT. Special attention was given to links. We provide both the hash of the full link and the hash of the top level domain.

On top of this, we also hashed user- and Tweet identifiers, preventing researchers from simply looking up the Tweet text via the Twitter API.
While we recognize that de-anonymization and hence the full reconstitution of the dataset is possible (and likely), we take this measure as a means to discourage misuse of the dataset. It is worth noting again that the original dataset was public already, hence any de-anonymization would not reveal any private information. 

\subsubsection{User sampling}

We took some extra precautions to make sure that the set of users that have positive engagements is similar to the set of users that had negative/no engagements. This is to give researchers more insight into specific users' histories. Separate uniform sampling for positives and negatives at Twitter's scale would have led to mostly disjoint user sets.

\subsection{Metrics}

In the following the two metrics used to evaluate the performance of a model are presented.

\subsubsection{Relative Cross Entropy}
Relative Cross Entropy (RCE) corresponds to the improvement of a prediction relative to the straw man, or the naive prediction, measured in cross entropy (CE). The naive prediction corresponds to the case that does not take into account the user and Tweet features, e.g., it always predicts the average (observed) CTR of the training set. Suppose the average CE of the naive prediction is $CE_{naive}$ and average CE of the prediction to be evaluated is $CE_{pred}$, then RCE is defined as $(CE_{naive} - CE_{pred}) \times 100/CE_{naive}$. Note that the lower the CE the better the quality of the predictions, so the higher the RCE. The benefit of using RCE is that we can obtain a confident estimate of whether the model is under or over performing the naive prediction.

\subsubsection{Area Under the Precision-Recall Curve (PR-AUC)}
Recall is equivalent to the true positive rate (or sensitivity) in a classification problem, while precision is the same as positive predictive value. Reviewing both precision and recall is particularly useful in cases there is an imbalance in the observations between two classes. The Area Under the Precision-Recall Curve (PR-AUC) is a commonly used evaluation metric and is more sensitive than AUC on skewed data.

\subsection{Baseline}
In this section, we will describe a simple baseline model architecture that works with the provided data format. It mainly constitutes of the following feature embedding and prediction components. 

\subsubsection{Numeric Features}
 For a numeric feature $num_i$ (e.g., follower count), we compute $n_q$ quantiles based on its cumulative distribution function and create $n_q + 1$ buckets, $(-\infty, q_i^1), ..., [-q_i^j, q_i^{j+1}), ... , [-q_i^{n_q}, +\infty)$, where $q_i^j$ denotes the $j^{th}$ quantile. Note that we also reserve an extra bucket for missing feature values. The feature $num_i$ is then bucketized into a one-hot encoded representation $e_i \in \mathbb{R}^{n_q+2}$, where $e_i^j = 1$ if $num_i$ falls into the $j^{th}$ bucket. 

\subsubsection{Categorical Features}
For categorical features $c_i$ (e.g., Tweet language), we one-hot encode them in $ \mathbb{R}^{N_{c_i}+1}$  where $N_{c_i}$ denotes the cardinally of the vocabulary. We reserve one extra dimension for out-of-vocabulary item. Binary features are considered a type of categorical features.

\subsubsection{Id Features}
For ID features $id_i$, the vocabulary is either unknown or of extremely large cardinality (e.g. user or Tweet id), we choose to hash $id_i$ into number, and then mod it into a given number of buckets.

\subsubsection{Tweet Text}
The text of the Tweet is tokenized and embedded using Chaos Free Recurrent Neural Network (CFRNN)~\cite{cfrnn}, chosen for computational efficiency and stability. In the datset we release the list of integers $(s_1, ..., s_l)$ corresponding to the index of the token in the embedding.

\subsubsection{Model}
For Numeric, Categorical, and ID Features, the corresponding one-hot encodings are converted to dense representations of size 16, then  concatenated along with Tweet feature embedding (embedding size 16). The obtained feature vector is then fed into a 3 layer multi-layer perceptron (hidden state size is 128, 64 and 32, activation function is leaky ReLU) to do the final predictions, in which the output size of the model is 4, corresponding to 4 engagement classes (Like, Reply, Retweet, and Retweet with comment). Since these four types of engagement are not mutually exclusive we use a sigmoid rather than softmax as the activation function in the final layer. 

For the baseline, $n_q$ is set to 49, resulting in 50 buckets in total. We use the huber loss~\cite{huber} for each class and the model is trained with Adam optimizer~\cite{adam} and learning rate 0.001 for 1 million steps. 

\section{Conclusion and Future Directions}\label{conc}
In this paper we have provided an overview of a rather challenging task: predicting user engagement with Tweets. To this end, a detailed and formal problem definition is provided, a set of concrete, tangible challenges faced within a real-world environment, and a set of the state-of-the-art approaches to motivate and further explore the task. We described the RecSys Challenge and the publicly released large-scale dataset, being an invitation for the broader research community to tackle this task. Finally, we also provided details of an exemplary, baseline approach developed upon the described dataset.

As we briefly addressed in the paper, various challenges emerge when delivering content such as Tweets to a user, but surely, there are numerous to be further explored. For instance, contextual information can enrich the recommendation model in order to deliver more appropriate content according to the contextual situation of a user at hand \cite{adomavicius2011context}, which was also shown in \cite{demaio2019} by simply accounting for the time of day when the user is active. Furthermore, we mentioned that engagement is used as a proxy for content quality, but this does not have to be the case, a user could engage with a certain content even when they dislike it (this even being the reason for engaging with it). Certainly, the models are tuned to accurately predict the next Tweet a user is likely to engage with, but other issues should as well be considered, such as, serendipity, diversity, coverage, etc., with a goal to truly comprehend the relevance of the delivered content. To tackle these and many other challenges, to broaden the knowledge-base, it is crucial that the data describing such user-content interactions is available for practitioners and researchers. Therefore, this paper is only a step forward to making a stronger cooperation between industry and academia.


%
%


%
%
\bibliographystyle{spphys} 
\bibliography{biblio}

\begin{thebibliography}{10}
\providecommand{\url}[1]{{#1}}
\providecommand{\urlprefix}{URL }
\expandafter\ifx\csname urlstyle\endcsname\relax
  \providecommand{\doi}[1]{DOI \discretionary{}{}{}#1}\else
  \providecommand{\doi}{DOI \discretionary{}{}{}\begingroup
  \urlstyle{rm}\Url}\fi

\bibitem{murthy2018}
D.~Murthy, \emph{Twitter} (Polity Press, Cambridge, UK, 2018)

\bibitem{bollen2011}
J.~Bollen, H.~Mao, X.~Zeng, Journal of computational science \textbf{2}(1), 1
  (2011)

\bibitem{weng2011}
J.~Weng, B.S. Lee, in \emph{Fifth international AAAI conference on weblogs and
  social media} (AAAI Press, CA, USA, 2011), pp. 401--408

\bibitem{aramaki2011}
E.~Aramaki, S.~Maskawa, M.~Morita, in \emph{Proceedings of the Conference on
  Empirical Methods in Natural Language Processing} (Association for
  Computational Linguistics, USA, 2011), EMNLP ’11, p. 1568–1576

\bibitem{coppersmith2014}
G.~Coppersmith, M.~Dredze, C.~Harman, in \emph{Proceedings of the workshop on
  computational linguistics and clinical psychology: From linguistic signal to
  clinical reality} (Association for Computational Linguistics, USA, 2014), pp.
  51--60.
\newblock \doi{10.3115/v1/W14-3207}

\bibitem{odea2015}
B.~O'dea, S.~Wan, P.J. Batterham, A.L. Calear, C.~Paris, H.~Christensen,
  Internet Interventions \textbf{2}(2), 183 (2015)

\bibitem{duan2010}
Y.~Duan, L.~Jiang, T.~Qin, M.~Zhou, H.Y. Shum, in \emph{Proceedings of the 23rd
  International Conference on Computational Linguistics} (Association for
  Computational Linguistics, USA, 2010), COLING ’10, p. 295–303

\bibitem{chen2012}
K.~Chen, T.~Chen, G.~Zheng, O.~Jin, E.~Yao, Y.~Yu, in \emph{Proceedings of the
  35th International ACM SIGIR Conference on Research and Development in
  Information Retrieval} (Association for Computing Machinery, New York, NY,
  USA, 2012), SIGIR ’12, p. 661–670.
\newblock \doi{10.1145/2348283.2348372}.
\newblock \urlprefix\url{https://doi.org/10.1145/2348283.2348372}

\bibitem{demaio2019}
C.~De~Maio, G.~Fenza, M.~Gallo, V.~Loia, M.~Parente, Future Generation Computer
  Systems \textbf{93}, 924 (2019)

\bibitem{pennacchiotti2012}
M.~Pennacchiotti, F.~Silvestri, H.~Vahabi, R.~Venturini, in \emph{Proceedings
  of the 21st ACM International Conference on Information and Knowledge
  Management} (Association for Computing Machinery, New York, NY, USA, 2012),
  CIKM ’12, p. 165–174.
\newblock \doi{10.1145/2396761.2396786}.
\newblock \urlprefix\url{https://doi.org/10.1145/2396761.2396786}

\bibitem{Knijnenburg2012Explaining}
B.P. Knijnenburg, M.C. Willemsen, Z.~Gantner, H.~Soncu, C.~Newell, User
  Modeling and User-Adapted Interaction \textbf{22}(4), 441 (2012).
\newblock \doi{10.1007/s11257-011-9118-4}.
\newblock \urlprefix\url{https://doi.org/10.1007/s11257-011-9118-4}

\bibitem{liu2009learning}
T.Y. Liu, et~al., Foundations and Trends{\textregistered} in Information
  Retrieval \textbf{3}(3), 225 (2009)

\bibitem{monti2017geometric}
F.~Monti, M.~Bronstein, X.~Bresson, in \emph{Proceedings of the 31st
  International Conference on Neural Information Processing Systems} (Curran
  Associates Inc., Red Hook, NY, USA, 2017), NIPS’17, p. 3700–3710

\bibitem{mikolov2013distributed}
T.~Mikolov, I.~Sutskever, K.~Chen, G.~Corrado, J.~Dean, in \emph{Proceedings of
  the 26th International Conference on Neural Information Processing Systems -
  Volume 2} (Curran Associates Inc., Red Hook, NY, USA, 2013), NIPS’13, p.
  3111–3119

\bibitem{mikolov2013efficient}
T.~Mikolov, K.~Chen, G.~Corrado, J.~Dean.
\newblock Efficient estimation of word representations in vector space (2013)

\bibitem{mikolov2013linguistic}
T.~Mikolov, W.t. Yih, G.~Zweig, in \emph{Proceedings of the 2013 Conference of
  the North {A}merican Chapter of the Association for Computational
  Linguistics: Human Language Technologies} (Association for Computational
  Linguistics, Atlanta, Georgia, 2013), pp. 746--751.
\newblock \urlprefix\url{https://www.aclweb.org/anthology/N13-1090}

\bibitem{devlin2018bert}
J.~Devlin, M.W. Chang, K.~Lee, K.~Toutanova.
\newblock Bert: Pre-training of deep bidirectional transformers for language
  understanding (2018)

\bibitem{Shiebler2018FightingRA}
D.~Shiebler, L.~Belli, J.~Baxter, H.~Xiong, A.~Tayal.
\newblock Fighting redundancy and model decay with embeddings (2018)

\bibitem{he2017neural}
X.~He, L.~Liao, H.~Zhang, L.~Nie, X.~Hu, T.S. Chua, in \emph{Proceedings of the
  26th International Conference on World Wide Web} (International World Wide
  Web Conferences Steering Committee, Republic and Canton of Geneva, CHE,
  2017), WWW ’17, p. 173–182.
\newblock \doi{10.1145/3038912.3052569}.
\newblock \urlprefix\url{https://doi.org/10.1145/3038912.3052569}

\bibitem{cheng2016wide}
H.T. Cheng, L.~Koc, J.~Harmsen, T.~Shaked, T.~Chandra, H.~Aradhye, G.~Anderson,
  G.~Corrado, W.~Chai, M.~Ispir, R.~Anil, Z.~Haque, L.~Hong, V.~Jain, X.~Liu,
  H.~Shah, in \emph{Proceedings of the 1st Workshop on Deep Learning for
  Recommender Systems} (Association for Computing Machinery, New York, NY, USA,
  2016), DLRS 2016, p. 7–10.
\newblock \doi{10.1145/2988450.2988454}.
\newblock \urlprefix\url{https://doi.org/10.1145/2988450.2988454}

\bibitem{guo2017deepfm}
H.~Guo, R.~TANG, Y.~Ye, Z.~Li, X.~He, in \emph{Proceedings of the Twenty-Sixth
  International Joint Conference on Artificial Intelligence} (International
  Joint Conferences on Artificial Intelligence Organization, CA, USA, 2017),
  pp. 1725--1731.
\newblock \doi{10.24963/ijcai.2017/239}

\bibitem{wang2017deep}
R.~Wang, B.~Fu, G.~Fu, M.~Wang, in \emph{Proceedings of the ADKDD’17}
  (Association for Computing Machinery, New York, NY, USA, 2017), ADKDD’17.
\newblock \doi{10.1145/3124749.3124754}.
\newblock \urlprefix\url{https://doi.org/10.1145/3124749.3124754}

\bibitem{naumov2019deep}
M.~Naumov, D.~Mudigere, H.J.M. Shi, J.~Huang, N.~Sundaraman, J.~Park, X.~Wang,
  U.~Gupta, C.J. Wu, A.G. Azzolini, et~al.
\newblock Deep learning recommendation model for personalization and
  recommendation systems (2019)

\bibitem{lian2018xdeepfm}
J.~Lian, X.~Zhou, F.~Zhang, Z.~Chen, X.~Xie, G.~Sun, in \emph{Proceedings of
  the 24th ACM SIGKDD International Conference on Knowledge Discovery \& Data
  Mining} (Association for Computing Machinery, New York, NY, USA, 2018), KDD
  ’18, p. 1754–1763.
\newblock \doi{10.1145/3219819.3220023}

\bibitem{koren2009matrix}
Y.~Koren, R.~Bell, C.~Volinsky, Computer \textbf{42}(8), 30 (2009)

\bibitem{chapelle2015simple}
O.~Chapelle, E.~Manavoglu, R.~Rosales, ACM Transactions on Intelligent Systems
  and Technology (TIST) \textbf{5}(4), 61 (2015)

\bibitem{bloom1970space}
B.H. Bloom, Communications of the ACM \textbf{13}(7), 422 (1970)

\bibitem{Sweeney1997GuaranteeingAW}
L.~Sweeney, in \emph{Proceedings: a conference of the American Medical
  Informatics Association. AMIA Fall Symposium} (Hanley \& Belfus, Inc.,
  Nashville, TN, USA, 1997), p. 51—55.
\newblock \urlprefix\url{https://europepmc.org/articles/PMC2233452}

\bibitem{Narayanan2006HowTB}
A.~Narayanan, V.~Shmatikov.
\newblock How to break anonymity of the netflix prize dataset (2006)

\bibitem{cfrnn}
T.~Laurent, J.~Brecht.
\newblock A recurrent neural network without chaos (2016)

\bibitem{huber}
P.J. Huber, The Annals of Mathematical Statistics \textbf{35}(1), 73 (1964)

\bibitem{adam}
D.~Kingma, J.~Ba.
\newblock Adam: A method for stochastic optimization (2014).
\newblock \urlprefix\url{http://arxiv.org/abs/1412.6980}

\bibitem{adomavicius2011context}
G.~Adomavicius, A.~Tuzhilin, in \emph{Recommender systems handbook} (Springer,
  Boston, MA, 2011), pp. 217--253

\end{thebibliography}

\end{document}